\documentclass{elsarticle}
\usepackage {graphicx}
\def\cjaa{Chinese J. Astron. Astrophys.}

\def\prb{Phys.~Rev.~B}
\def\pre{Phys.~Rev.~E}
\def\jcp{Journal of Chemical  Physics} 

\begin {document}
\begin {frontmatter}

\title {
Poissonian and non Poissonian Voronoi Diagrams
with application 
to the aggregation of molecules
}
\author {L. Zaninetti}
\address {Dipartimento di Fisica Generale,
 via P.Giuria 1,\\ I-10125 Turin,Italy }
\ead {zaninetti@ph.unito.it}
\ead [url]{http://www.ph.unito.it/$\tilde{~}$zaninett}

\begin {abstract}
The distributions that regulate  the spatial domains
of the Poissonian Voronoi Diagrams are discussed
adopting the sum of gamma variate of argument two.
The distributions that arise from the 
product and quotient of two
gamma variates of argument two are also derived.
Three examples of non Poissonian seeds for the Voronoi Diagrams
are discussed.
The developed algorithm allows the 
simulation of an aggregation of methanol and water.
\end {abstract}
\begin {keyword}
 89.75.Kd; 02.50.Ng; 02.10.-v;
Voronoi diagrams; Monte Carlo methods;
Cell-size distribution
\end {keyword}
\end {frontmatter}

\section{Introduction}

A great number of natural phenomena are
described by Poissonian Voronoi diagrams , 
we cite some of them :
lattices in quantum field theory \cite{Drouffe1984} ;
 conductivity and
percolation in granular composites  \cite{Jerauld1984_a,Jerauld1984_b};
 modeling growth of metal clusters on amorphous
substrates \cite{Dicenzo1989};
the statistical mechanics of simple glass 
forming systems in $2D$ \cite{Hentschel2007};
modeling of material interface evolution 
in grain growth of polycrystalline
materials \cite{Lee2006}.
We now outline some applications in 
Chemistry.
Detailed computer simulation results of 
several static and dynamical
properties of water, obtained by 
using a realistic potential model 
were analyzed 
by evaluating the volume distributions of 
Voronoi polyhedron as well as
angular and radial distributions of molecular clusters 
\cite{Ruocco1993}.
The local structure of three hydrogen bonded liquids 
comprising clusters of
markedly different topology: water, methanol, and HF are investigated by
analyzing the properties of the Voronoi polyhedron 
of the molecules in
configurations obtained from Monte Carlo computer simulations
 \cite{Jedlovszky2000}.
The local lateral structure of 
dimyristoylphosphatidylcholine
cholesterol mixed membranes of different compositions has been
investigated on the basis 
of the Voronoi polygons  \cite{Jedlovszky2004}.
Molecular dynamics simulation of a linear 
soft polymer has been performed
and the free volume properties of the system 
have been analyzed in detail in
terms of the Voronoi polyhedron of the monomers
\cite{Sega2004}.
The molecular dynamics simulation of the aqueous solutions 
of urea of
different concentrations 
are modeled by the 
method of Voronoi polyhedron  \cite{Idrissi2008}.

From the point of view of the theory
the only known analytical result for the
Poissonian Voronoi diagrams , in the following $V-P$
after the two memories \cite{voronoi_1907,voronoi}
 ,
is the size distribution in $1D$ .
More precisely the theoretical 
distribution of segments in $1D$ 
is the gamma variate distribution of argument
two , in the following $GV2$  \cite{kiang}.
The area in $2D$ and the volume in $3D$ were conjectured to follow
the sum of two and three $GV2$ \cite{kiang}.
The possibility 
that the segments , area , and volume of $V-P$
can be modeled by a unique formula parametrized 
with $d$ which represents the considered dimension 
$d(d=1,2,3)$ has been recently analyzed~\cite{Ferenc_2007}.
Section~\ref{secsum} of this paper 
reviews 
the known formulas
of  the sum of $GV2$ 
and express them according to the chosen dimension.
Section~\ref{secproduct} explores the product and quotient
of two $GV2$.
Section~\ref{secnon} analyzes three examples 
of Voronoi Diagrams with correlated seeds or 
non Poissonian processes.
Section~\ref{secchimica} reports the simulation
of an 
aggregation of methanol and water.

\section{Sum of a Gamma variate }
\label{secsum}

\begin{figure}
\begin{center}
\includegraphics[width=10cm]{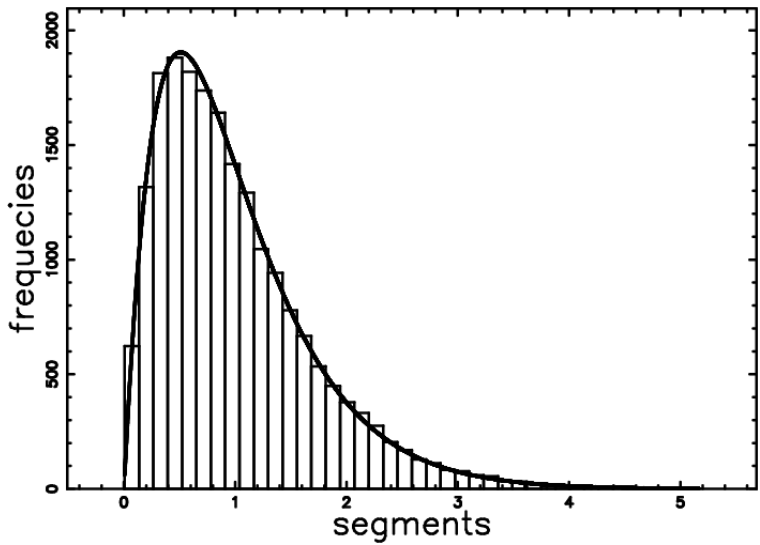}
\end {center}
\caption { Histogram (step-diagram) of the $V-P$
normalized segment distribution in $1D$ with a superposition of the
gamma--variate as represented by equation~(\ref{kiangd}):
the number of Poissonian seeds is 20000 ,
$d=1.02$ , 
$c=2.04$ ,
$NBIN=40$ 
 and $\chi^2$=32.52 ,random seeds}
 \label{gamma1d}%
 \end{figure}

The starting point is the
probability density function ( in the following PDF)
 in length , $s$ ,
of a segment in a random fragmentation
\begin{equation}
p(s) = \lambda \exp {(-\lambda s)} ds
\quad ,
\end{equation}
where $\lambda$ is the hazard rate of 
the exponential distribution.
Given the fact that the sum , $u$ , of two exponential
distributions has a  PDF
\begin{equation}
p(u)= \lambda^2 u \exp{(- \lambda u)} du
\quad .
\end{equation}
The PDF of the $1D$ $V-P$ segments , $l$,
( the midpoint of the sum of two segments) can be found
in  the previous formula inserting $u=2l$
\begin{equation}
p(l) = 2 \lambda l \exp{(-2 \lambda l)} d (2 \lambda l)
\quad .
\end{equation}
When  transformed  in normalized units $x=\frac{l}{\lambda}$
the following PDF is obtained 
\begin{equation}
p(x) = 2 x \exp{(-2 x) } d (2 x)
\quad .
\end{equation}
When this result is expressed as a gamma variate
we obtain the PDF (formula~(5) in \cite{kiang})
\begin{equation}
 H (x ;c ) = \frac {c} {\Gamma (c)} (cx )^{c-1} \exp(-cx)
\quad ,
\label{kiang}
\end{equation}
where $ 0 \leq x < \infty $ , $ c~>0$
and $\Gamma (c)$ is the gamma function with argument c;
in the case of $1D$ $V-P$ $c=2$.
As an example  Figure~\ref{gamma1d} reports 
the histogram
of length of the normalized Voronoi segments in 1D.

The Kiang  PDF has a mean of 
\begin{equation}
\mu = 1
\quad ,
\end{equation}
and variance
\begin{equation}
\sigma^2 = \frac{1}{c}
\quad .
\end{equation}
It was conjectured that the area in $2D$ and the volumes
in $3D$ of the $V-P$ may be approximated
as the sum of two and three $GV2$ .
Due to the fact that the sum of n independent gamma
variates with shape parameter $c_i$ is a gamma variate with
shape parameter $c = \sum_{i}^{n} c_i$,
the area and  volumes are supposed to follow a gamma variate
with argument 4 and 6 ~\cite{Feller_1971,Tribelsky2002}.
This hypothesis was later named
 "Kiang  conjecture", and equation (\ref{kiang}) was used as
a fitting function \cite{Zaninetti1992,kumar,Zaninetti2006,
Korobov2009},
or as a  hypothesis to accept or to reject using the standard
procedures of the data analysis,
see~\cite{Tanemura1988,Tanemura2003}.
The PDF (\ref{kiang}) can be generalized by
introducing the
dimension of the considered space, $d(d=1,2,3)$
\cite{Tanemura2003}, \cite{Hinde1980} 
\begin{equation}
 H (x ;d ) = \frac {2d} {\Gamma (2d)} (2dx )^{2d-1} \exp(-2dx)
\quad .
\label{kiangd}
\end{equation}
Two other PDFs give interesting results in 
the operation of fit of
the $V-P$ area/volume. The first is the generalized gamma
PDF with three parameters $(a,b,c)$ , 
\cite{Hinde1980,Ferenc_2007,Tanemura2003} , 
\begin{equation}
f(x;b,c,d) = c \frac {b^{a/c}} {\Gamma (a/c) } x^{a-1} \exp{(-b
x^c)} \quad . \label{gammag}
\end{equation}
The generalized gamma
PDF has the mean of 
\begin{equation}
\mu = \frac
{
{b}^{-\frac{1}{c} }\Gamma \left( \frac {1+a}{c} \right) 
}
{
\Gamma \left( {\frac {a}{c}} \right) 
}
\quad ,
\end{equation}
and variance
\begin{equation}
\sigma^2 = \frac
{
{b}^{- \frac{2}{c} } \left( +\Gamma \left( {\frac {2+a}{c}} \right) 
\Gamma \left( {\frac {a}{c}} \right) - \left( \Gamma \left( {\frac {
1+a}{c}} \right) \right) ^{2} \right) 
}
{
\left( \Gamma \left( {\frac {a}{c}} \right) \right) ^{2}
}
\quad .
\end{equation}
By the method of maximum likelihood the estimators ,
$\widehat{a}$, $\widehat{b}$ and $\widehat{c}$
are the solution to the simultaneous equations
\begin{eqnarray}
\frac {n}{\widehat c} ln (\widehat b)
-n \frac {\Psi(\frac{\widehat a}{\widehat c})}{\widehat c}
+ \sum_{i=1}^n \ln( x_i) =0 \nonumber \\
n \frac{\widehat a} {\widehat c} \frac{1}{\widehat b} 
- \sum_{i=1}^n \; x_i^{\widehat c} = 0 \\
 \frac{n}{\widehat c} 
-n \frac{\widehat a}{{ \widehat c}^2} ln ( \widehat b) 
+n \frac {\Psi(\frac{ \widehat a}{ \widehat c}) \widehat a }
{{ \widehat c}^2}
- { \widehat b} \sum_{i=1}^n \; x_i^{\widehat c} ln (x_i) 
=0 \nonumber 
\quad ,
\end {eqnarray} 
where $\Psi (x)$ is the digamma function ,
$n$ the number of observations in a sample
and $x_i$ is an observed value. 
The second one is a
PDF of the type \cite{Ferenc_2007}
\begin{equation}
FN(x;d) = Const \times x^{\frac {3d-1}{2} } \exp{(-(3d+1)x/2)}
\quad ,
\label{rumeni}
\end{equation}
where
\begin{equation}
Const =
\frac
{
\sqrt {2}\sqrt {3\,d+1}
}
{
2\,{2}^{3/2\,d} \left( 3\,d+1 \right) ^{-3/2\,d}\Gamma \left( 3/2\,d+
1/2 \right)
}
\quad ,
\end{equation}
and $d(d=1,2,3)$ represents the
dimension of the considered space.
We will call the previously reported 
function the  Ferenc-Neda  PDF  
which has  the mean  of 
\begin{equation}
\mu = 1
\quad ,
\end{equation}
and variance
\begin{equation}
\sigma^2 = \frac{2}{3d+1}
\quad .
\end{equation}

\begin{figure}
\begin{center}
\includegraphics[width=10cm]{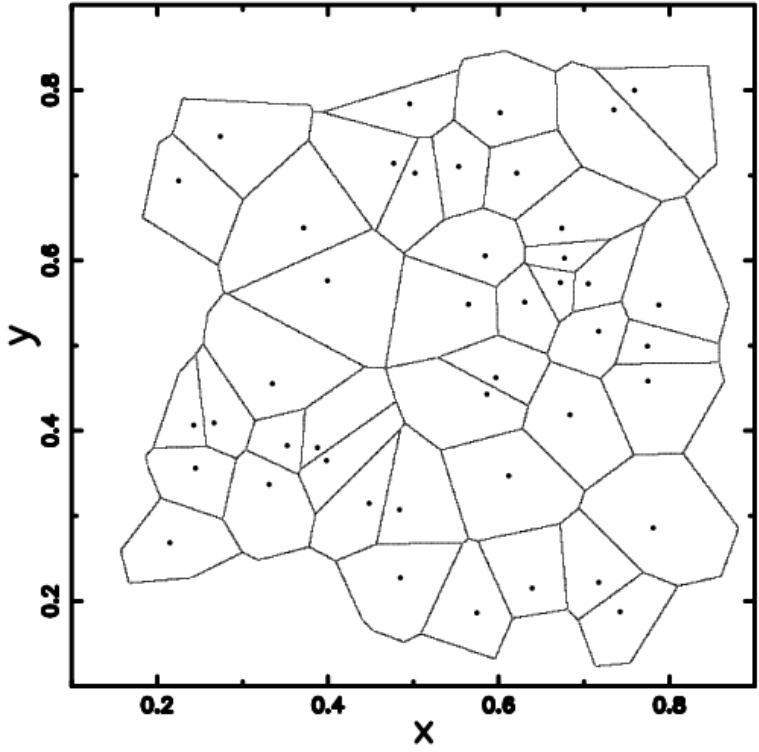}
\end {center}
\caption { The $V-P$ in 2D.
The selected region comprises 44 random seeds marked by a
point. } \label{voro_network}
\end{figure}

Which distribution produces the best fit of the area of the
irregular polygons and the volume of the irregular Polyhedron ? 
In
order to answer this question we fitted the sample of the area
and volume with the three distributions here considered . 
The 2D
$V-P$ are reported in Figure~\ref{voro_network} ,
 Table~\ref{data}
reports the $\chi^2$ results
and Table~\ref{area_abc} reports the parameters 
of generalized gamma deduced in two papers 
with the addition 
of  our values.

 \begin{table}
 \caption[]{Values of $\chi^2$  for 
the cells normalized area-distribution function in 2D.
Here we have 
25087 Poissonian seeds and $40$ intervals in the histogram; 
$\nu$ denotes the degrees of freedom.}
 \label{data}
 \[
 \begin{array}{lll}
 \hline
PDF ~& \nu ~& \chi^2  \\ \noalign{\smallskip}
 \hline
 \noalign{\smallskip}
k(x ;d) ~(Eq. (\ref{kiangd}))~,d=1.77                  &  39 & 83.48   \\
\noalign{\smallskip} f(x;d) ~(Eq.(\ref{rumeni}))~      &  39 & 71.83  \\
\noalign{\smallskip} G(x;a,b,c) ~ (Eq. (\ref{gammag})) &  36 & 58.9 \\ 
\noalign{\smallskip} h(x)  ~ (Eq. (\ref{bessel}))      &  39 & 12636 \\ 
\noalign{\smallskip} r(x)  ~ (Eq. (\ref{ratio}))       &  39 & 19986 \\ 
 \hline
 \hline
 \end{array}
 \]
 \end {table}

\begin{figure}
\begin{center}
\includegraphics[width=10cm]{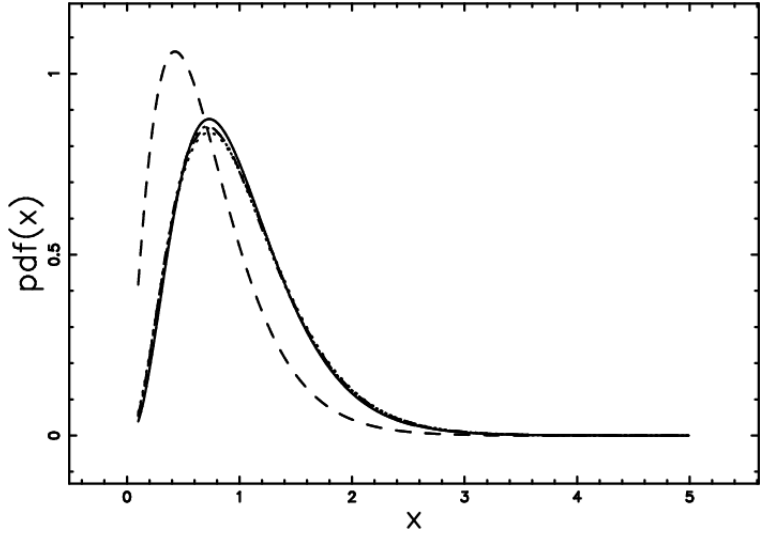}
\end {center}
\caption
{ 
Poissonian seeds.
Plot of 
the cells normalized area-distribution function in $2D$
when 5 PDFs are adopted :
k(x ;d) ~(Eq. (\ref{kiangd}))~,d=1.87
 (full line) ,
 f(x;b,c,d) with data of Ferenc-Neda~2007 
 (dashed),
 f(x;b,c,d) with data of Tanemura~1988~and~2003
 (dot-dash-dot-dash) ,
 f(x;b,c,d) with our data 
 (dotted)
and
f(x;d) ~(Eq.(\ref{rumeni})) ,d=2,
(dash-dot-dot-dot) .
}
 \label{vedi_gamma_area}%
 \end{figure}

\begin{table}
 \caption[]{Generalized gamma parameters $a$, $b$ and $c$ 
of the normalized area distribution in $2D$ in different papers
with Poissonian seeds.
}
 \label{area_abc}
 \[
 \begin{array}{lccc}
 \hline
reference  & a & b & c \\ \noalign{\smallskip}
 \hline
 \noalign{\smallskip}
Ferenc-Neda~2007       & 2.29 & 3.01 & 1.08 \\
Tanemura~1988~and~2003 & 3.31 & 3.04 & 1.078 \\
this~paper             & 3.15 & 2.72 & 1.13 \\
\\ \noalign{\smallskip}
 \hline
 \hline
 \end{array}
 \]
 \end {table}
Figure~\ref{vedi_gamma_area} reports the various 
PDFs here adopted for 
the normalized area distribution in $2D$.

\begin{figure}
\begin{center}
\includegraphics[width=10cm]{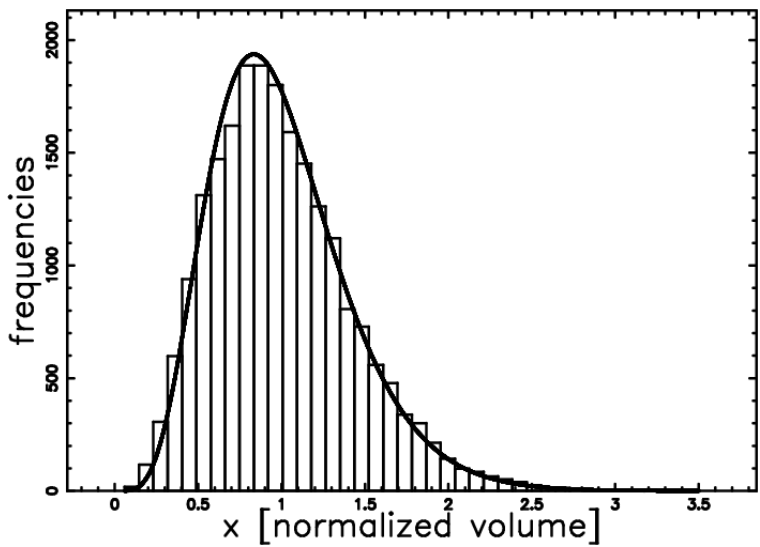}
\end {center}
\caption { Histogram (step-diagram) of the $V-P$
normalized volume distribution in $3D$ with a superposition of the
gamma--variate as represented by equation~(\ref{kiangd}):
the number of Poissonian seeds is 1392 ,
d=3,
NBIN=40 and $\chi^2$=62.63 ,random seeds}
 \label{gamma_kiang_6}%
 \end{figure}

In $3D$ Figure~\ref{gamma_kiang_6} reports the histogram of the
volume distribution as well as the Kiang's PDF~\ref{kiangd} when
$d=3$ ,
Table~\ref{datavolume} reports the $\chi^2$ results
and Table~\ref{volume_abc} reports the parameters 
of generalized gamma in two papers with the addition
of  our values.

 \begin{table}
 \caption[]{Values of $\chi^2$  
for distribution of normalized volumes of $3D$ cells. 
Here 21378 Poissonian seeds were generated and 
$40$ interval were used in the histogram. }
\label{datavolume}
 \[
 \begin{array}{llll}
 \hline
PDF ~& \nu ~& \chi^2  \\ \noalign{\smallskip}
 \hline
 \noalign{\smallskip}
k(x ;d) ~(Eq. (\ref{kiangd}))~,d=2.76                   & 39 & 93.86  \\
\noalign{\smallskip} f(x;d) ~(Eq.(\ref{rumeni}),d=3)~       & 39 & 134.15  \\
\noalign{\smallskip} G(x;a,b,c) ~ (Eq. (\ref{gammag}))  & 36 & 58.59  \\ 
 \hline
 \hline
 \end{array}
 \]
 \end {table}
Table~\ref{volume_abc} reports the parameters 
of generalized gamma PDF deduced in 
two papers with the addition of our values.

\begin{table}
 \caption[]{Generalized gamma parameters $a$ , $b$ and $c $
of the normalized volume distribution in $3D$ 
in different papers with Possonian seeds .}
 \label{volume_abc}
 \[
 \begin{array}{lcccc}
 \hline
reference  & a & b & c \\ \noalign{\smallskip}
 \hline
 \noalign{\smallskip}
Ferenc-Neda~2007 & 3.24 & 3.24 & 1.26 \\
Tanemura~1988~and~2003 & 4.80 & 4.06 & 1.16 \\
this~paper & 4.68 & 3.87 & 1.18 \\
\\ \noalign{\smallskip}
 \hline
 \hline
 \end{array}
 \]
 \end {table}

\begin{figure}
\begin{center}
\includegraphics[width=10cm]{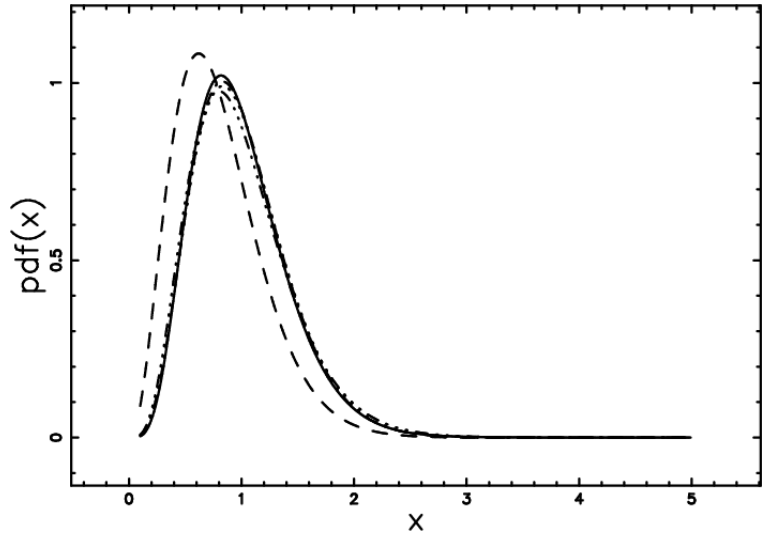}
\end {center}
\caption
{ 
Poissonian seeds.
Plot of 
the cells normalized volume-distribution function in $3D$
when 5 PDFs are adopted :
k(x ;d) ~(Eq. (\ref{kiangd}))~,$d= 2.79$ or $c=5.58$ 
 (full line) ,
 f(x;b,c,d) with data of Ferenc-Neda~2007 
 (dashed),
 f(x;b,c,d) with data of Tanemura~1988~and~2003
 (dot-dash-dot-dash) ,
 f(x;b,c,d) with our data 
 (dotted)
and
f(x;d) ~(Eq.(\ref{rumeni})) ,d=3,
(dash-dot-dot-dot) .
}
 \label{vedi_gamma_volume}%
 \end{figure}

Figure~\ref{vedi_gamma_volume} reports the various 
PDFs here adopted for 
the normalized volume distribution in $3D$.

\section{Product and Quotient of a Gamma variate }
\label{secproduct}

This Section explores
the product XY and the quotient X/Y when
X and Y are $GV2$.

\subsection{Product}

We recall that if
 $X$ is a random variable of the continuous type with 
 PDF 
, $f(x)$,
 which
is defined and positive on the interval
 $ 0 \leq x < \infty $
and similarly if
$Y$
is a random variable of the continuous type with PDF
$g(y)$
 which is defined
and
positive $ 0 \leq y < \infty $ , the PDF of $V = XY$ is
\begin{equation}
h(v) =
\int_0^\infty g (\frac{v}{x}) f(x) \frac{1}{x} dx
\quad .
\end{equation}
Here the case of equal limits of integration will
be explored , when this is not true difficulties arise
\cite{Springer1979,Glen2004} .
When $f(x)$ and $g(y)$ are
$GV2$
the PDF is
\begin{equation}
h(v) =
\int_0^\infty
\frac {
16\,{e^{-2\,x}}v{e^{-2\,{\frac {v}{x}}}}
}
{
x
}
dx
=
32\,v{\it K_0} \left( 4\,\sqrt {v} \right)
\quad ,
\label{bessel}
\end{equation}
where $ K_{\nu}(z) $ is the modified Bessel function of
the second kind \cite{Abramowitz1965,press}
with $\nu$ representing the order, in our case 0.
The previous integral can be solved with the
substitution $x=\frac{y}{2}$ and using the integral representation
in \cite{Watson2008} , pag. 183
\begin{equation}
 K_{\nu}(z) = \frac{1}{2} (\frac{1}{2} z) ^{\nu}
 \int_0^\infty \exp \bigl (-\tau - \frac{z^2}{4\tau} \bigr )
 \frac {1}{\tau^{\nu+1}} d \tau
 \quad .
\end{equation}
The mean of the new PDF, $h(v)$, as represented
by formula~(\ref{bessel}) is
\begin{equation}
<v> =\int_0^\infty
v \times 32\,v{\it K_0} \left( 4\,\sqrt {v} \right) dv = 1
\quad .
\end{equation}
The mode , $m$ , is at v =0.15067 and 
Table~\ref{data} reports the $\chi^2$
of the fit of the $V-P$ normalized area-distribution in $2D$.

\subsection{The quotient}

The PDF of $V = X/Y$ 
when
X , represented by g(x) , 
and Y 
, represented by f(y) , 
are $GV2$
is
\begin{equation}
r(v) =
\int_0^\infty |w| g ( {v}{x}) f(x) dx
= \frac
{ 
6\,v
}
{
 \left( 1+v \right) ^{2} \left( 1+2\,v+{v}^{2} \right) 
}
\quad .
\label{ratio}
\end{equation}
The mean of the new PDF, $r(v)$, as represented
by formula~(\ref{ratio}) is
\begin{equation}
<v> =\int_0^\infty
v \times 
\frac
{ 
6\,v
}
{
 \left( 1+v \right) ^{2} \left( 1+2\,v+{v}^{2} \right) 
}
 dv = 2
\quad .
\end{equation}
The mode , $m$ , is at $v =\frac{1}{3}$
and 
Table~\ref{data} reports the $\chi^2$
of the fit of the Voronoi cell's normalized 
area-distribution in 2D.

\section{Non-Poissonian seeds}

\label{secnon}
We now explore the case in which the seeds of the 
Voronoi Diagrams are distributed in a correlated 
way with  respect to the center of the box.
The correlated seeds are generated in polar coordinates ,
$\rho$ and $\theta$.
The radius $\rho$ , the distance from the center of the box ,
is generated according to the PDF which is the product of 
two $GV2$ , see formula~(\ref{bessel})
once the scale parameter $b$ is introduced ,
\begin{equation}
h(v,b) = \frac{
32\,v{\it K_0} \left( 4\,\sqrt {\frac{v}{b}} \right)
}
{b^2}
\quad .
\label{besselb}
\end{equation}
Random numbers  of the polar angle $\theta$ , in degrees,
 can be generated 
from those of the unit rectangular variate $R$ using the relationship
\begin{equation}
\theta \sim 360 \times R 
\quad .
\end{equation}

\begin{figure}
\begin{center}
\includegraphics[width=10cm]{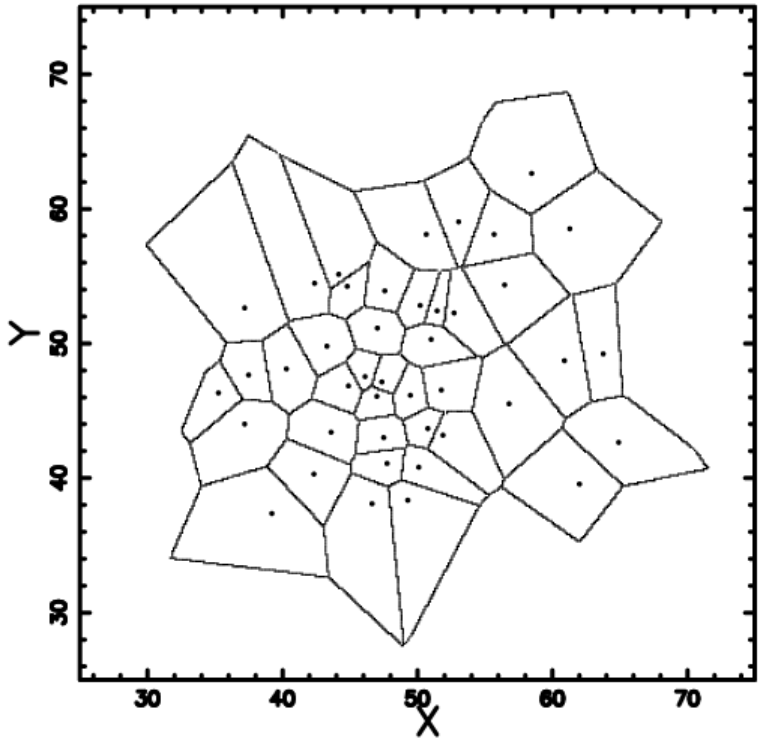}
\end {center}
\caption { The Voronoi Diagram in 2D
with Non-Poissonian seeds generated
according to 
the product 
of two $GV2$ , 
see formula~(\ref{besselb}). 
The selected region comprises 42 random seeds marked by a
point
and
$b=25$.
 } \label{network_altre}
\end{figure}

A typical example of Voronoi Diagrams generated
by such  seeds is reported in Figure~\ref{network_altre} 
and Table~(\ref{datak}) 
reports the $\chi^2$ of three different fits.
From a careful analysis of Table~(\ref{datak}) 
it is possible to conclude that the product of 
two random variables,
produces a better fit with  respect to the Kiang function
with fixed or variable $d$.

\begin{table}
 \caption[]{Normalized area distribution in $2D$
 of 166 correlated seeds generated according to 
formula~(\ref{besselb}).
The $\chi^2$ of data fit when the number
 of classes is 40 for three PDF }
 \label{datak}
 \[
 \begin{array}{lll}
 \hline
PDF & name & \chi^2 \\ \noalign{\smallskip}
 \hline
 \noalign{\smallskip}
H (x ;d ) & Kiang~PDF ~,formula~(\ref{kiangd})~when~ d=2~, & 2142 \\
 \noalign{\smallskip}
H (x ;d ) & Kiang~PDF ~,formula~(\ref{kiangd})~when~ d=0.61~, & 43 \\
\noalign{\smallskip}h(v) & product~two~gamma, formula~(\ref{bessel}) & 40 \\
\noalign{\smallskip}
 \hline
 \hline
 \end{array}
 \]
 \end {table}

Another example is represented 
by the ratio of two $GV2$ , formula~(\ref{ratio}) ,
that when the scale parameter $b$ is introduced
has PDF 
\begin{equation}
r(v,b) = 
\frac
{
6\,v{b}^{4}
}
{
{b}^{2} \left( b+v \right) ^{2} \left( {b}^{2}+2\,vb+{v}^{2}
 \right) 
}
\quad .
\label{quotientb}
\end {equation}

\begin{figure}
\begin{center}
\includegraphics[width=10cm]{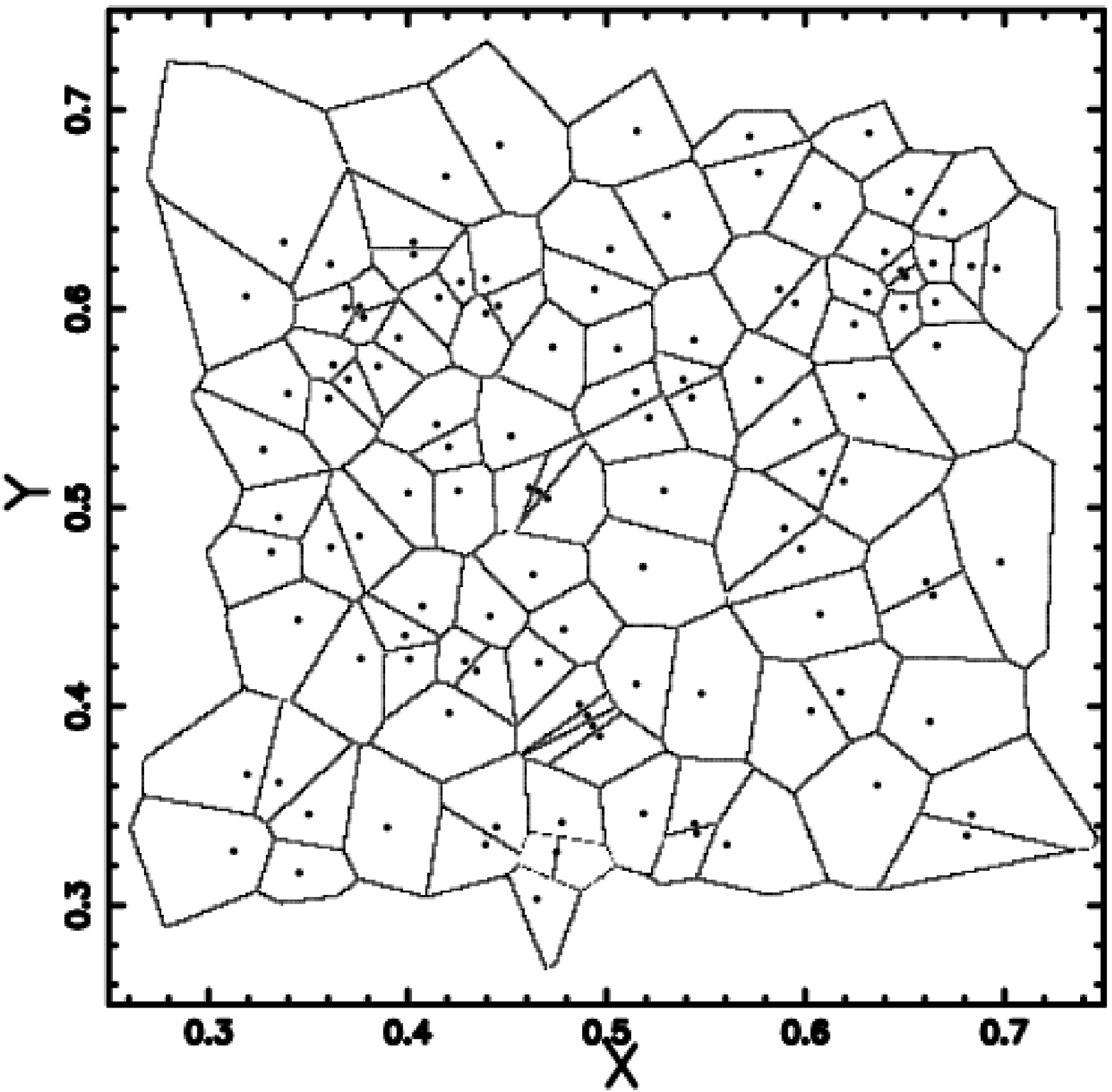}
\end {center}
\caption { 
The Voronoi--diagram in $2D$ when 
the seeds are generated according to 
the quotient 
of two $GV2$ , 
see formula~(\ref{quotientb}) .
The selected region comprises 117 seeds and
$b=25$.
 } 
\label{area_rapp}
\end{figure}

An example of Voronoi Diagrams
generated by seeds that follow 
the quotient 
of two $GV2$ is reported in Figure~\ref{area_rapp} 
and Table~\ref{tablequotient}
reports the $\chi^2$ of three different fits.
\begin{table}
 \caption[]{Normalized area distribution in $2D$ 
 of 117 correlated seeds generated according to 
formula~(\ref{quotientb}).
The $\chi^2$ of data fit when the number
 of classes is 40 for three PDF }
 \label{tablequotient}
 \[
 \begin{array}{lll}
 \hline
PDF & name & \chi^2 \\ \noalign{\smallskip}
 \hline
 \noalign{\smallskip}
H (x ;d ) & Kiang~PDF ~,formula~(\ref{kiangd})~when~ d=2~, & 5560 \\
 \noalign{\smallskip}
H (x ;d ) & Kiang~PDF ~,formula~(\ref{kiangd})~when~ d=1~, & 87 \\
\noalign{\smallskip}r(v) & ratio~two~gamma, formula~(\ref{ratio}) & 79 \\
\noalign{\smallskip}
 \hline
 \hline
 \end{array}
 \]
 \end {table}

From a careful analysis of Table~\ref{tablequotient} 
it is possible to conclude that the quotient of two $GV2$ ,
also in this case produces a better fit 
with respect to the Kiang function
with fixed or variable $d$.

A third example is represented 
by the Normal ( Gaussian ) distribution 
which  has PDF 
\begin{equation}
N(x;\sigma) = 
\frac {1} {\sigma (2 \pi)^{1/2}} \exp ({- {\frac {x^2}{2\sigma^2}}} )
\quad 
-\infty < x < \infty 
\quad .
\label{gaussian}
\end{equation}
When only positive values of $x$ are considered PDF~(\ref{gaussian})
transforms in 
\begin{equation}
HN(x;\sigma) = 
\frac {2} {\sigma (2 \pi)^{1/2}} \exp ({- {\frac {x^2}{2\sigma^2}}} )
\quad 
0 < x < \infty 
\quad .
\label{gaussianhalf}
\end{equation}

We now consider
the product of two normal random variables
$X = N(0,1)$ and $Y = N(0,1)$.
The PDF of $V=XY$ is \cite{Glen2004} 
\begin {equation}
 hn(v) = \left \{ \begin {array} {ll}
 K_0(v * signum (v))/ \pi & -\infty < v < 0 \\
 K_0(v * signum (v))/ \pi & 0 < v < \infty ~.
\end {array}
 \right .
\end {equation}
This PDF has a pole at $v=0$ and 
when the scale parameter $b$ is introduced 
and only positive values are considered 
it transforms in
\begin {equation}
 hn(v;b) = 2 K_0(\frac{v}{b} )/ \pi \quad 0 < v < \infty 
 \quad .
\label{productgauss}
\end {equation}

\begin{figure}
\begin{center}
\includegraphics[width=10cm]{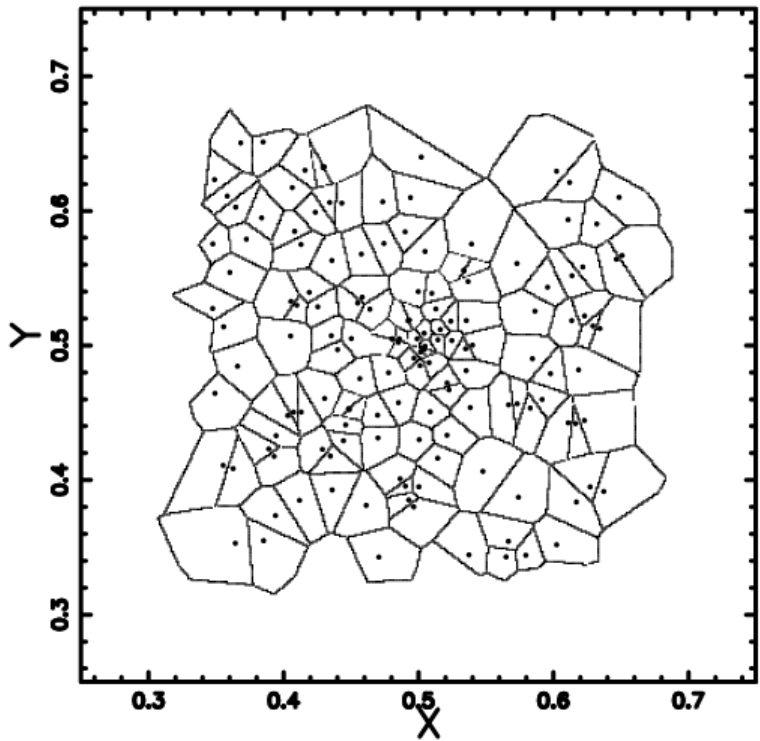}
\end {center}
\caption { 
The Voronoi--diagram in $2D$ when 
the seeds are generated according to 
the Half Normal (Gaussian) distribution 
,
see formula~(\ref{gaussianhalf}) .
The selected region comprises 145 seeds ,$side=1$ and 
$\sigma=0.5$.
 } 
\label{area_gauss}
\end{figure}

An example of Voronoi Diagrams
generated by seeds that follow 
the Half Normal (Gaussian) distribution (formula(\ref{gaussianhalf})) 
is reported in Figure~\ref{area_gauss} 
and Table~\ref{tablegauss}
reports the $\chi^2$ of four different fits.

\begin{table}
 \caption[]{Normalized area distribution in $2D$
 of 145 correlated seeds generated according to 
formula~(\ref{gaussianhalf}), {side}=1, $\sigma=0.5$.
The $\chi^2$ of data fit when the number
 of classes is 40 for three PDF }
 \label{tablegauss}
 \[
 \begin{array}{lll}
 \hline
PDF & name & \chi^2 \\ \noalign{\smallskip}
 \hline
 \noalign{\smallskip}
H (x ;d ) & Kiang~PDF ~,formula~(\ref{kiangd})~when~ d=2~, & 2070 \\
 \noalign{\smallskip}
H (x ;d ) & Kiang~PDF ~,formula~(\ref{kiangd})~when~ d=0.88~, & 84 \\
\noalign{\smallskip}
HN(x;\sigma) & Half~ Normal(Gaussian)~PDF , formula~(\ref{gaussianhalf}) &
168 \\
\noalign{\smallskip}
hn(v;b) & Product ~two~Gaussian~PDF , formula~(\ref{productgauss}) & 
78 \\
\noalign{\smallskip}
 \hline
 \hline
 \end{array}
 \]
 \end {table}
From a careful analysis of Table~\ref{tablegauss} it 
is possible to conclude that the 
product of two Gaussian PDFs
produces a better fit with  respect to the 
Half Normal (Gaussian) distribution and the
Kiang function.

\section{Aggregation of molecules}
\label{secchimica}

\begin{figure}
\begin{center}
\includegraphics[width=10cm]{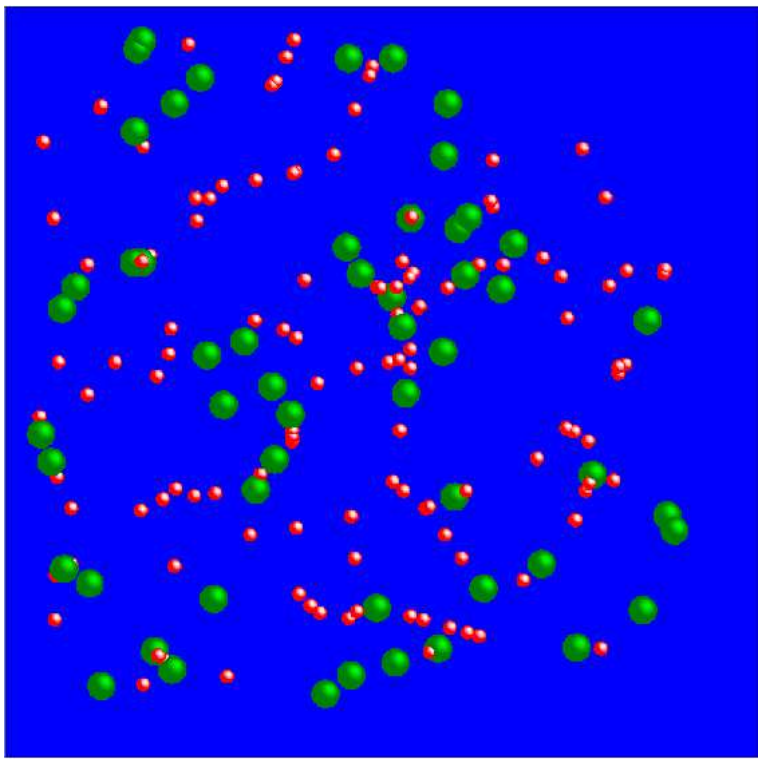}
\end {center}
\caption 
{ 
Superposed on the $V-P$ network of Figure~\ref{voro_network}
are present 131 molecules of water (red small balls)
and 54 molecules of methanol (green great balls).
} 
\label{molecole}
\end{figure}

The analysis of the molecular dynamics 
through the Voronoi polyhedron in $3D$ or
the Voronoi polygons in $2D$ is becoming a 
standard procedure.
What is 
very interesting is the surface aggregation 
of a 
water-methanol mixture at 298 $K$,
see~\cite{Partay2008}.
The aggregation of the methanol and water
molecules at the surface of their mixture
is analyzed in  the light of the Voronoi
diagrams considering   water  and methanol molecules
together ,
see Figure~11 in~\cite{Partay2008}.
Figure~12 in~\cite{Partay2008}
reports an
instantaneous snapshot of the surface layer 
of a system
containing 5 $\%$
 methanol molecules (top view), as taken out from an
equilibrium configuration.
In order to simulate such an aggregate 
Figure~\ref{voro_network} reports the $2D$ 
$V-P$ and Figure~\ref{molecole} 
the theoretical displacement 
of the molecules on such a network.

\begin{figure}
\begin{center}
\includegraphics[width=10cm]{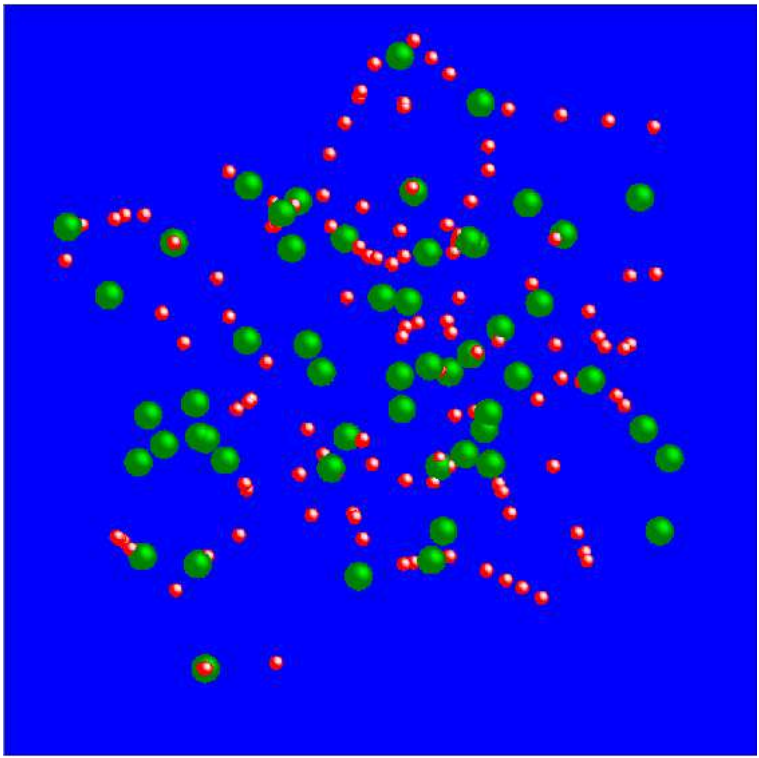}
\end {center}
\caption 
{ 
Superposed on the network of Figure~\ref{network_altre}
generated by non Poissonian seeds 
are present 131 molecules of water (red small balls)
and 54 molecules of methanol (green great balls).
} 
\label{molecole_altre}
\end{figure}

The same aggregate is simulated 
by non Poissonian seeds
and Figure~\ref{network_altre} reports the
Voronoi Diagram 
while  Figure~\ref{molecole_altre} 
the theoretical displacement 
of the molecules on such correlated network.

\section{Conclusions}

The size distribution of $V-P$ in $1D$ 
 has an exact analytical result
that is a $GV2$.
Starting from this analytical result we derived the sum ,
the product and the quotient of two $GV2$.
The area distribution of $V-P$ in $2D$ 
is well represented by the sum rather than the product of
and quotient of two $GV2$ , see Table~\ref{data}.
A careful comparison between the Kiang  function,
formula (\ref{kiangd}) ,
and a new function suggested by 
Ferenc-Neda , see formula~(\ref{rumeni}),
has been done , see Table~\ref{data} and Table~\ref{datavolume}.
In presence of non-Poissonian seeds
that present a symmetry around the center
the situation is inverted
and the product of two $GV2$ ,
formula~(\ref{bessel}),
or the ratio of two $GV2$
,formula~(\ref{ratio}) , 
produce a better fit of the 
area distribution of Voronoi polygons in $2D$ 
with respect 
to the Kiang function, formula (\ref{kiangd}) ,
see Table~\ref{datak} and Table~\ref{tablequotient}.
A third test made on Normal (Gaussian) correlated 
seeds is in agreement with the conjecture
that the area of the $2D$ Voronoi diagrams follows the
distribution of the seeds.
The developed algorithm allows us to simulate 
some well studied aggregation such as the molecules 
of the water-methanol mixture.
The question of whether if the area of the voids between molecules 
follows  the  $V-P$  area distribution 
as represented by the Kiang function , see 
equation~(\ref{kiang}) or 
the gamma of Ferenc-Neda  , see equation~(\ref{rumeni})
is open  to future efforts.

\end{document}